# Large effective mass and interaction-enhanced Zeeman splitting of $K$-valley electrons in MoSe$_2$

Stefano Larentis,[1] Hema C. P. Movva,[1] Babak Fallahazad,[1] Kyounghwan Kim,[1] Armand Behroozi,[1] Takashi Taniguchi,[2] Kenji Watanabe,[2] Sanjay K. Banerjee,[1] and Emanuel Tutuc[1, *]

[1]*Microelectronics Research Center, Department of Electrical and Computer Engineering,
The University of Texas at Austin, Austin, TX 78758, USA*
[2]*National Institute for Materials Science, 1-1 Namiki Tsukuba, Ibaraki 305-0044, Japan*

(Dated: May 16, 2018)

We study the magnetotransport of high-mobility electrons in monolayer and bilayer MoSe$_2$, which show Shubnikov-de Haas (SdH) oscillations and quantum Hall states in high magnetic fields. An electron effective mass of $0.8m_e$ is extracted from the SdH oscillations' temperature dependence; $m_e$ is the bare electron mass. At a fixed electron density the longitudinal resistance shows minima at filling factors (FFs) that are either predominantly odd, or predominantly even, with a parity that changes as the density is tuned. The SdH oscillations are insensitive to an in-plane magnetic field, consistent with an out-of-plane spin orientation of electrons at the $K$-point. We attribute the FFs parity transitions to an interaction enhancement of the Zeeman energy as the density is reduced, resulting in an increased Zeeman-to-cyclotron energy ratio.

Group VI transition metal dichalcogenides (TMDs) 1H-monolayers are direct bandgap two-dimensional (2D) semiconductors with band extrema at the corners ($K$-point) of the hexagonal Brillouin zone [1]. The combination of strong spin-orbit interaction (SOI) and broken inversion symmetry results in a large bandgap at the $K$-point, and a spin-split bandstructure with coupled spin and valley degrees of freedom [2–4]. Magnetotransport in clean TMD samples can be used to probe the energy-momentum dependence at the band extrema, the Landau level (LL) structure, and assess the impact of electron-electron interaction via negative compressibility or enhanced Zeeman splitting. Shubnikov-de Haas (SdH) oscillations of $K$-valley holes in mono- and bilayer WSe$_2$ have revealed predominantly two-fold degenerate LLs [5], and interaction-enhanced Zeeman splitting [6, 7]. Similarly, Γ-valley holes in few-layer WSe$_2$ show large effective masses and enhanced Zeeman splitting [8]. Magnetotransport of 2D electrons in TMDs has been hindered by challenges in obtaining high-mobility samples and low-temperature Ohmic contacts [9]. Magnetotransport in few-layer MoS$_2$ and WS$_2$ samples reveal three or six-fold degenerate LLs, consistent with $Q$-valley conduction band (CB) extrema [10–12]. Compressibility studies of monolayer WSe$_2$ reveal comparable $K$-valley electron and hole effective masses, and interaction-enhanced LL Zeeman splitting in the valence band (VB), but not in the CB [7].

Here we report a study of SdH oscillations in high-mobility electrons in dual-gated mono- and bilayer MoSe$_2$, using Pd bottom-contacts. From the temperature dependence of the SdH oscillations amplitude, we extract an electron effective mass of $0.8m_e$; $m_e$ is the bare electron mass. We observe predominantly even or odd filling factors (FFs) depending on the electron density ($n$), an observation explained by an interaction-enhanced Zeeman splitting with reducing density. Tilted magnetic-field measurements indicate that the electron spin is locked perpendicular to the MoSe$_2$ plane.

Our devices are fabricated using MoSe$_2$ flakes exfoliated from synthetic crystals (HQ Graphene). Mono- and bilayer flakes are identified using a combination of Raman and pho-

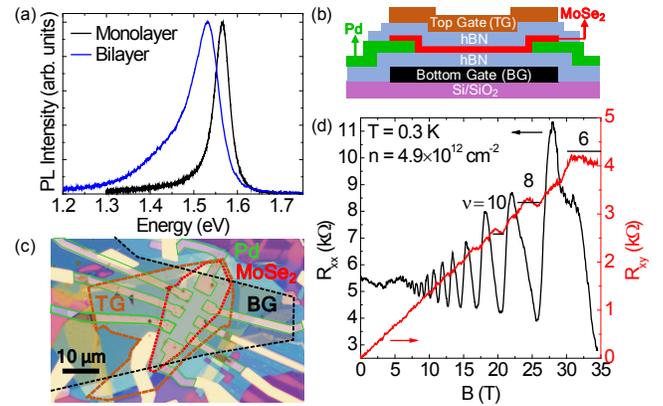

FIG. 1. (a) Normalized room temperature PL spectra of mono- and bilayer MoSe$_2$. (b) Schematic cross-section and (c) optical micrograph of a dual-gated, hBN-encapsulated MoSe$_2$ device. Outlines of different colors mark the MoSe$_2$ flake (red), Pd contacts (green), top (orange) and bottom (black) graphite gates. (d) $R_{xx}$ (left axis) and $R_{xy}$ (right axis) vs $B$ measured at $T$ = 0.3 K and $n$ = $4.9 \times 10^{12}$ cm$^{-2}$ in bilayer MoSe$_2$ B1.

toluminescence (PL) spectroscopy. Figure 1(a) shows the normalized PL spectra for both mono- and bilayer flakes, at room temperature, using an excitation wavelength of 532 nm. The monolayer (bilayer) PL spectrum features a single prominent peak at 1.57 (1.53) eV, associated with the A exciton [13, 14]. Figure 1(b) shows a cross-section schematic of a dual-gated, hexagonal boron nitride (hBN)-encapsulated MoSe$_2$ device with bottom Pd contacts, fabricated using a layer pick-up and transfer method [15, 16]. Figure 1(c) shows an optical micrograph of a device with top and bottom graphite gates. Devices with metal gates show similar results. The Pd bottom contacts along with MoSe$_2$ electrostatic doping at positive top-gate bias ($V_{TG}$) provide $n$-type Ohmic contacts at low-temperatures. Data from two monolayer(A1, A2), and three bilayer (B1, B2, B3) MoSe$_2$ samples are included in this study. The measurements were carried out at temperatures down to $T$ = 0.3 K,

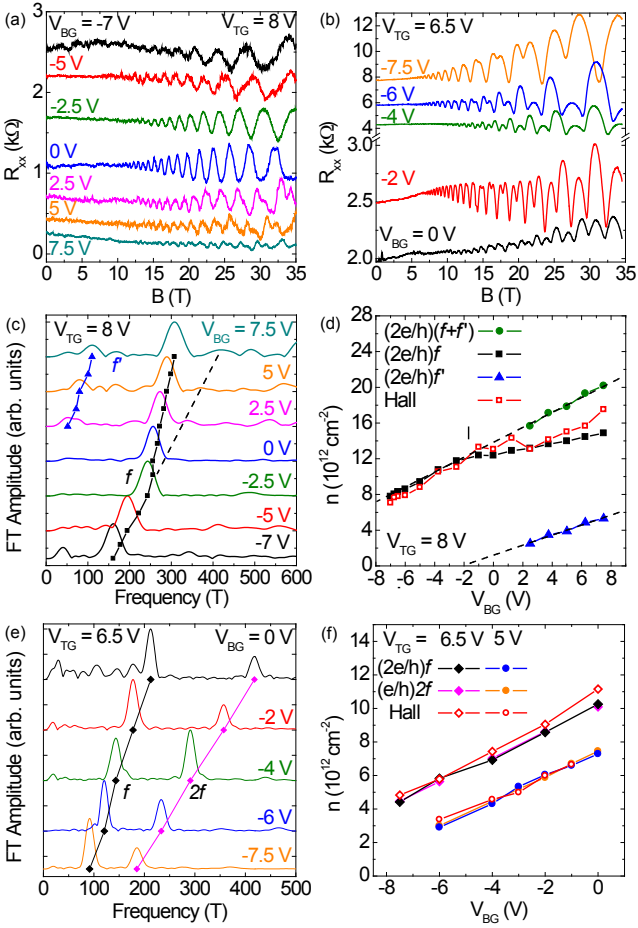
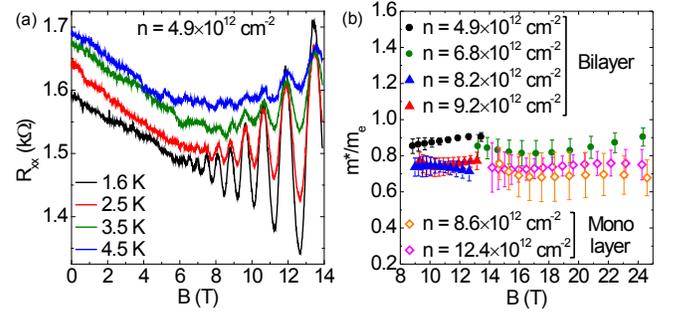

FIG. 2. (a) $R_{xx}$ vs $B$ measured at various $V_{BG}$ values, $V_{TG} = 8$ V, and $T = 0.3$ K in monolayer MoSe$_2$ A1. (b) $R_{xx}$ vs $B$ measured at various $V_{BG}$ values, $V_{TG} = 6.5$ V, and $T = 1.5$ K in bilayer MoSe$_2$ B2. The traces in panels (a,b) are offset for clarity. (c),(e) Normalized FT amplitude vs frequency corresponding to $R_{xx}$ vs $B^{-1}$ data of panel (a) and (b), respectively. (d) $n$ vs $V_{BG}$ measured in monolayer MoSe$_2$ A1 at $V_{TG} = 8$ V. The onset of the upper spin-split subband population is marked. (f) $n$ vs $V_{BG}$ measured in bilayer MoSe$_2$ B2 at $V_{TG} = 6.5$ V (diamonds) and $V_{TG} = 5$ V (circles). Solid (open) symbols correspond to $n$ determined from FT ($R_{xy}$) data.

FIG. 3. (a) $R_{xx}$ vs $B$ measured at various $T$ values, at $n = 4.9 \times 10^{12}$ cm$^{-2}$ in bilayer MoSe$_2$ B1. (b) $m^*/m_e$ vs $B$ measured at different $n$ in monolayer MoSe$_2$ A1 ($\diamond$), bilayer MoSe$_2$ B1 ($\bullet$), and B2 ($\blacktriangle$).

and magnetic fields up to 35 T.

Figure 1(d) shows the longitudinal ($R_{xx}$) and Hall ($R_{xy}$) resistance as a function of the perpendicular magnetic field ($B$) measured in bilayer MoSe$_2$ sample B1 at $n = 4.9 \times 10^{12}$ cm$^{-2}$, and $T = 0.3$ K. The data show SdH oscillations developing at $B > 6$ T, corresponding to a mobility $\mu \simeq 1650$ cm$^2$/Vs. At high $B$-fields quantum Hall states (QHSs) develop at $\nu = 6, 8, 10$; $\nu = nh/eB$, where $e$ is the electron charge, and $h$ is Planck's constant. Similar data measured in monolayer MoSe$_2$ sample A1 are included in the Supplemental Material [17].

Figures 2(a,b) show $R_{xx}$ vs $B$ measured at different bottom-gate biases ($V_{BG}$), in monolayer A1 at $V_{TG} = 8$ V, $T = 0.3$ K and in bilayer B2 at $V_{TG} = 6.5$ V, $T = 1.5$ K, respectively. Figures 2(c) and 2(e) show The Fourier transform (FT) amplitude vs frequency corresponding to $R_{xx}$ vs $B^{-1}$ data of Fig. 2(a) and 2(b), respectively. The FT is performed by first subtracting a polynomial background from the $R_{xx}$ vs $B^{-1}$ data to center it around zero, followed by a Hamming window multiplication, and a fast FT algorithm.

Figure 2(c) data, corresponding to monolayer MoSe$_2$, reveal one principal peak at a frequency ($f$) for $V_{BG} \leq 0$ V. For $V_{BG} > 0$ V, $f$ shows a weaker $V_{BG}$ dependence, and a second, lower frequency peak ($f'$) emerges, indicating a second subband is populated. The subband, $(2e/h)f$ and $(2e/h)f'$, and the total $(2e/h)(f + f')$ densities, along with the $n$ values determined from the $R_{xy}$ slope at low $B$-fields are summarized as a function of $V_{BG}$ in Fig. 2(d). The electron density determined from the SdH oscillation frequency is obtained assuming two-fold degenerate LLs. The total $n$ displays a linear dependence on $V_{BG}$. At $n > 12.5 \times 10^{12}$ cm$^{-2}$ the second subband ($f'$) is populated, as marked in Fig. 2(d). The SOI leads to a splitting of the spin-up and spin-down states at the $K$-point in TMDs. This splitting is $\approx 0.2$ eV and $\approx 25$ meV for monolayer MoSe$_2$ VB [1] and CB [4, 18, 19], respectively. We associate the peaks $f$ and $f'$ in Fig. 2(c-d) with the population of the lower and upper CB spin-split bands of monolayer MoSe$_2$, respectively.

Figure 2(e) data, corresponding to bilayer MoSe$_2$, reveal one principal peak at a frequency $f$, and its second harmonic ($2f$) indicating a single subband is occupied. The $f$ value increases linearly with $V_{BG}$, consistent with Fig. 2(c) data in monolayer MoSe$_2$ with only the lowest spin-split subband populated. Figure 2(f) shows a comparison between $n = (2e/h)f$ calculated using the $f$ values of Fig. 2(e), and the $n$ values determined from the $R_{xy}$ slope at low $B$-fields as a function of $V_{BG}$.

Figure 3(a) shows $R_{xx}$ vs $B$ data measured at various $T$ values, at constant $n = 4.9 \times 10^{12}$ cm$^{-2}$ in bilayer B1. Using the temperature ($T$) dependence of the SdH oscillations amplitude ($\Delta R_{xx}$) we extract the electron effective mass ($m^*$), as $\Delta R_{xx} \propto \xi/\sinh\xi$, where $\xi = 2\pi^2 k_B T/\hbar\omega_c$ and $\omega_c = eB/m^*$; $k_B$ is the Boltzmann constant, and $\hbar$ is the reduced Planck's constant [17]. Figure 3(b) shows $m^*/m_e$ vs $B$ data for monolayer A1, and bilayer B1, B2 at $n$ ranging between $4.9 - 12.4 \times 10^{12}$ cm$^{-2}$, where only the lower spin-split CB at the $K$-point is probed.

The average $m^*/m_e = 0.8$ is largely insensitive to $n$ and $B$. Theoretical calculations of $m^*/m_e$ in monolayer MoSe$_2$ range between $0.50 - 0.56$ [4, 19, 20]. The measured $m^*$ values, and the corresponding density of states ($m^*/\pi\hbar^2$) allows us to determine the CB spin-splitting ($2\Delta_{cb}$) in monolayer MoSe$_2$. Considering the threshold density for the population of the upper CB subband $n_T = 12.5 \times 10^{12}$ cm$^{-2}$ [Fig. 2(d)], we obtain $2\Delta_{cb} = n_T \cdot \pi\hbar^2/m^* = 37$ meV, a value comparable to, albeit larger than theoretical calculations [4, 18, 19].

The CB minima are expected to be at the $K$-point in monolayer, and at the $Q$-point in bulk MoSe$_2$ [21, 22]. The data of Figs. 1-3 allow us to unambiguously determine the CB minima in mono- and bilayer MoSe$_2$. The two-fold LL degeneracy observed in both mono- and bilayer samples is consistent with CB minima at the $K$-point, as SdH oscillations of carriers at the $Q$-point show three- or six-fold degenerate LLs [10, 11]. The similar $m^*$ values of Fig. 3(b) for mono- and bilayer MoSe$_2$ further support this conclusion. In group VI TMD bilayers, the weak inter-layer coupling of $K$-valley carriers leads to two distinct subbands for each layer [5], with densities that can be independently controlled by $V_{TG}$ and $V_{BG}$. For $V_{TG} > 0$ V and $V_{BG} \leq 0$ V only the top layer is populated, and the bilayer MoSe$_2$ can be effectively treated as a monolayer. The absence of a beating pattern in bilayer SdH oscillations up to $n = 11.0 \times 10^{12}$ cm$^{-2}$ [Fig. 2(b)] indicates the electrons populate the lower spin-split subband of the top layer.

Figure 4(a) shows $R_{xx}$ vs $\nu$ at different $n$ values between $2.9 - 11.0 \times 10^{12}$ cm$^{-2}$ measured in bilayer B2. For $n$ values larger than $8.6 \times 10^{12}$ cm$^{-2}$, $R_{xx}$ minima are present at predominantly odd FFs. At $n = 7.0 \times 10^{12}$ cm$^{-2}$, the $R_{xx}$ minima at odd and even FFs are of equal strength up to $\nu = 36$. As $n$ is lowered to $5.6 \times 10^{12}$ cm$^{-2}$ the FF sequence turns predominantly even, and at $n = 4.5 \times 10^{12}$ cm$^{-2}$ the odd FFs $R_{xx}$ minima are absent. At the lowest $n = 2.9 \times 10^{12}$ cm$^{-2}$ another transition to odd FFs is observed. We note that at fixed $n$ the FF sequence is insensitive to changes in the transverse electric field [17].

To better understand the $n$-dependent FF sequence, we write the LLs CB energies $E_{l,\tau s} = \tau s\Delta_{cb} + (l + 1/2)E_c + sg_s\mu_B B/2 + \tau g_v\mu_B B/2$, where $l = 0, 1, 2, \ldots$ is the LL orbital index, $s = \pm 1$ corresponds to the electron spin $\uparrow$ and $\downarrow$, $\tau = \pm 1$ to the $K$ and $K'$ valleys, $E_c = \hbar\omega_c$ is the cyclotron energy, $\mu_B$ is the Bohr magneton, and $g_v$, $g_s$ are the valley and spin $g$-factors, respectively. The $\tau s\Delta_{cb}$ term describes the spin-split CB minima where the LLs originate. The $\tau s = \pm 1$ doublets lead to two LL fan diagrams with an energy separation of $2\Delta_{cb}$ at $B = 0$. We assume that electrons reside in the lowest spin-split band ($\tau s = -1$), where the total, spin and valley LL Zeeman energy is $E_Z|_{\tau s=-1} = \tau g^*\mu_B B$; $g^* = g_v - g_s$ is the effective $g$-factor for LLs of the lowest CB spin-split subband. The LL energies of the $\tau s = -1$ group write: $E_{l,\tau} = (l + 1/2)E_c + \tau g^*\mu_B B/2$. We use here the single-band model convention in which all LLs are two-fold degenerate in absence of Zeeman splitting [4, 23]. Using a model in which the $l = 0$ is non-degenerate [3] is equivalent

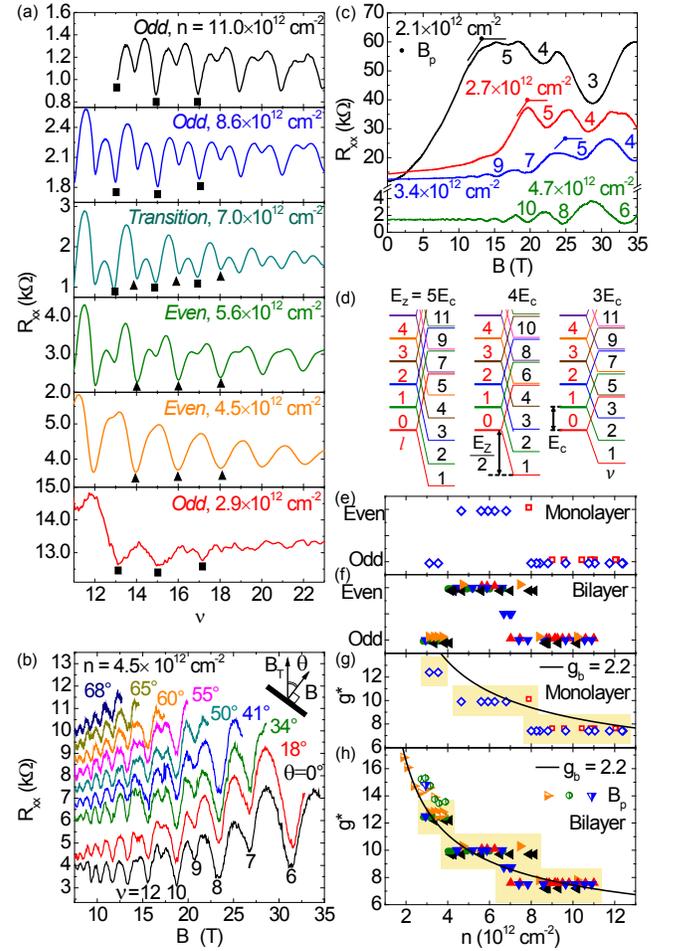

FIG. 4. (a) $R_{xx}$ vs $\nu$ measured at $n$ values between $2.9 - 11.0 \times 10^{12}$ cm$^{-2}$, $T = 1.5$ K in bilayer MoSe$_2$ B2. The FF sequence undergoes parity transitions at $n = 7.0 \times 10^{12}$ cm$^{-2}$, and $n = 4.0 \times 10^{12}$ cm$^{-2}$. The triangles (squares) mark $R_{xx}$ minima at even (odd) FFs. (b) $R_{xx}$ vs $B$ measured at different $\theta$, at $n = 4.5 \times 10^{12}$ cm$^{-2}$, and $T = 1.5$ K in bilayer B2. The traces are offset for clarity. Inset: sample orientation schematic. (c) $R_{xx}$ vs $B$ measured at $n$ between $2.1 - 4.7 \times 10^{12}$ cm$^{-2}$, $T = 0.3$ K in bilayer B3. (d) LLs structure highlighting the interplay between $E_Z$ and $E_c$. An even (odd) $E_Z/E_c$ corresponds an even (odd) FF sequence. (e,f) FF parity vs $n$ in mono- and bilayer MoSe$_2$ respectively. Symbol legend: monolayer A1 ($\diamond$), A2 ($\circ$); bilayer B1 ($\bullet$), B2 ($\blacktriangle,\blacktriangledown$), B3 ($\blacktriangleleft,\blacktriangleright$), $\blacktriangle,\blacktriangledown$ and $\blacktriangleleft,\blacktriangleright$ label different cooldowns. (g,h) $g^*$ vs $n$ in mono- and bilayer MoSe$_2$, respectively, and fit to the QMC calculations using $g_b = 2.2$ (solid line). The shaded region indicates the $g^*$ error bar $\Delta g^* = \pm m_e/m^*$.

to a $g^*$ offset by $2m_e/m^*$.

The Zeeman-to-cyclotron energy ratio determines the FF sequence, with even (odd) $E_Z/E_c$ values leading to even (odd) FFs. Figure 4(a) data reveal a $B$-field independent FF sequence at a fixed $n$, indicating that $E_Z/E_c$ does not vary with the $B$-field. The FFs parity transitions can be explained by an $n$-dependent $E_Z/E_c$, or equivalently by an $n$-dependent, interaction enhanced $g^*$. Consistent with the large effective mass, electron-electron interaction is expected to enhance $g^*$ as $n$ is reduced, as reported in Si [24, 25], GaAs [26], AlAs [27], and WSe$_2$ [6–8] 2D systems.

Magnetotransport in magnetic fields tilted at an angle ($\theta$) from the 2D plane normal [Fig. 4(b) inset] has been employed to probe the Zeeman splitting in 2D systems. If $E_Z$ is proportional to the total magnetic field ($B_T$) the FF sequence changes with $\theta$ [24]. Figure 4(b) shows $R_{xx}$ vs $B$ at various $\theta$ values and $n = 4.5 \times 10^{12}$ cm$^{-2}$ in bilayer B2. At $\theta = 0°$ the FF sequence is predominantly even, and remains unchanged for all $\theta$ values, indicating that $E_Z$ is insensitive to the parallel magnetic field component. These findings contrast observations in Si [24, 25], GaAs [26], AlAs [27], and few layer WSe$_2$ [8] 2D systems, but are in agreement with observations in tri-layer MoS$_2$ [11], and mono- and bilayer WSe$_2$ [6], where the combination of strong SOI and band extrema away from the Brillouin zone center locks the carrier spin perpendicular to the 2D system.

Figure 4(c) shows examples of $R_{xx}$ vs $B$ measured in bilayer B3 at low $n$ values. For $n < 4.0 \times 10^{12}$ cm$^{-2}$ the data show QHSs at consecutive FFs above a density-dependent field ($B_p$), where the occupied LLs have the same spin orientation. Interestingly, the observation of consecutive FFs above $B_p$ is accompanied by a pronounced positive magntoresistance (MR) background superimposed onto the SdH oscillations for $B < B_p$, similar to the positive MR associated with a parallel magnetic-field-induced spin polarization in Si, GaAs and AlAs 2D systems [25, 27, 28].

A quantitative determination of $g^*$ is possible using FF sequence parity data [Fig. 4(a)], and the spin-polarization field [Fig. 4(c)]. Figure 4(d) illustrates the LL structure, where the $E_c$ and $E_Z$ contributions are shown separately for different $E_Z/E_c$ values and FF sequences. Figures 4(e) and 4(f) summarize the FF sequence parity vs $n$ measured in mono- and bilayer samples respectively. Comparing Fig. 4(d) diagram and the FF sequence ($\nu$ = 4, 5, 7, 9, 11 . . .), associated to $R_{xx}$ vs $B$ data measured at $n = 3.4 \times 10^{12}$ cm$^{-2}$ in bilayer B3 [Fig. 4(c)], allows us to assign $E_Z/E_c = 5$ to the lowest $n$ FF parity group of Fig. 4(f). The observation of consecutive integer FFs above a certain magnetic field [Fig. 4(c)] allows to unambiguously assign $E_Z/E_c$. As $n$ is increased, each FF sequence transition is associated with a decrease in $E_Z$ equal to $E_c$ [Fig. 4(e,f)], consistent with a decreasing $g^*$ as the 2D system becomes less dilute. A FF sequence associated with a transition is assigned a half integer $E_Z/E_c$ value. Once we assign an $i = E_Z/E_c$ value to each FF sequence group [Figs. 4(e,f)], namely $i$ = 5, 4, 3, we determine $g^* = (2m_e/m^*)i$ as a function of $n$ as shown in Fig. 4(g,h) for both mono- and bilayer samples, respectively. At the onset of full spin polarization $E_Z$ is equal to the the Fermi energy, and $B_p = 2hn/(eg^*m^*/m_e)$ [28]. At low $n$ values the $B_p$ vs $n$ measurement provides a separate method to determine $g^*$ vs $n$. The $g^*$ values obtained from $B_p$ values and FF sequence transitions are summarized in Fig. 4(h) for bilayer samples.

Quantum Monte Carlo (QMC) spin susceptibility calculations [29] have shown good agreement with experiments in GaAs [26] and AlAs [27] 2D electrons, and in WSe$_2$ 2D holes in the $K$-valley [6]. A comparison between the measured $g^*$ and QMC results requires the band $g$-factor value ($g_b$) in absence of interaction effects. As the $g_b$ value remains to be established for MoSe$_2$ [4, 23, 30], we estimate $g_b$ = 2.2 using a fit of the QMC spin susceptibility [29] to the experimental $g^*$ vs $n$ data for both mono- [Fig. 4(g)] and bilayer [Fig. 4(h)] samples assuming implicitly the QMC calculations approximate well the interaction enhancement of $g^*$ in MoSe$_2$ as in other 2D systems [6, 26, 27]. The $n$ value is converted into a dimensionless inter-particle distance $r_s = 1/(\sqrt{\pi n}a_B^*)$, where $a_B^* = a_B(\kappa m_e/m^*)$ is the effective Bohr radius, and $\kappa$ the effective dielectric constant [31]; $a_B$ is the Bohr radius.

In summary, we report magnetotransport studies in high mobility mono- and bilayer MoSe$_2$. The SdH oscillations reveal a density dependent FF sequence, and a $K$-valley electron effective mass of $0.8m_e$. The FF sequence is insensitive to a parallel magnetic field, indicating the electron's spin is locked perpendicular to the MoSe$_2$ plane. The interplay between cyclotron and Zeeman energy, along with interaction enhanced, density dependent $g$-factor explains the FF sequence odd-to-even transitions. These findings clarify the LL structure of $K$-valley electrons in MoSe$_2$, and highlight the role of interactions in this large effective mass 2D system.

We thank D. Graf and A. Suslov for technical assistance, and X. Li and A. Kormányos for discussions. This work was supported by the Nanoelectronics Research Initiative SWAN center, Intel Corporation, and National Science Foundation Grant No. EECS-1610008. A portion of this work was performed at the National High Magnetic Field Laboratory, which is supported by National Science Foundation Cooperative Agreement No. DMR-1157490, DMR-1644779, and the State of Florida.

* etutuc@mer.utexas.edu
[1] D. Xiao, G.-B. Liu, W. Feng, X. Xu, and W. Yao, Phys. Rev. Lett. **108**, 196802 (2012).
[2] Z. Y. Zhu, Y. C. Cheng, and U. Schwingenschlögl, Phys. Rev. B **84**, 153402 (2011).
[3] X. Li, F. Zhang, and Q. Niu, Phys. Rev. Lett. **110**, 066803 (2013).
[4] A. Kormányos, V. Zólyomi, N. D. Drummond, and G. Burkard, Phys. Rev. X **4**, 011034 (2014).
[5] B. Fallahazad, H. C. P. Movva, K. Kim, S. Larentis, T. Taniguchi, K. Watanabe, S. K. Banerjee, and E. Tutuc, Phys. Rev. Lett. **116**, 086601 (2016).
[6] H. C. P. Movva, B. Fallahazad, K. Kim, S. Larentis, T. Taniguchi, K. Watanabe, S. K. Banerjee, and E. Tutuc, Phys. Rev. Lett. **118**, 247701 (2017).
[7] M. V. Gustafsson, M. Yankowitz, C. Forsythe, D. Rhodes, K. Watanabe, T. Taniguchi, J. Hone, X. Zhu, and C. R. Dean, Nature Materials **17**, 411 (2018).
[8] S. Xu *et al.*, Phys. Rev. Lett. **118**, 067702 (2017).
[9] X. Cui *et al.*, Nat. Nanotechnol. **10**, 534 (2015); Nano Lett. **17**, 4781 (2017).
[10] Z. Wu *et al.*, Nat. Commun. **7**, 12955 (2016).

- [11] R. Pisoni, Y. Lee, H. Overweg, M. Eich, P. Simonet, K. Watanabe, T. Taniguchi, R. Gorbachev, T. Ihn, and K. Ensslin, Nano Lett. **17**, 5008 (2017).
- [12] Q. H. Chen, J. M. Lu, L. Liang, O. Zheliuk, A. Ali, P. Sheng, and J. T. Ye, Phys. Rev. Lett. **119**, 147002 (2017).
- [13] P. Tonndorf *et al.*, Opt. Express **21**, 4908 (2013).
- [14] J. S. Ross *et al.*, Nat. Commun. **4**, 1474 (2013).
- [15] K. Kim *et al.*, Nano Lett. **16**, 1989 (2016).
- [16] S. Larentis, B. Fallahazad, H. C. P. Movva, K. Kim, A. Rai, T. Taniguchi, K. Watanabe, S. K. Banerjee, and E. Tutuc, ACS Nano **11**, 4832 (2017).
- [17] See Supplemental Material for additional magnetotransport data in monolayer MoSe$_2$, transverse electric-field dependence, and details about the effective mass extraction.
- [18] K. Kośmider, J. W. González, and J. Fernández-Rossier, Phys. Rev. B **88**, 245436 (2013).
- [19] A. Kormányos, G. Burkard, M. Gmitra, J. Fabian, V. Zólyomi, N. D. Drummond, and V. Falko, 2D Mater. **2**, 022001 (2015).
- [20] D. Wickramaratne, F. Zahid, and R. K. Lake, J. Chem. Phys. **140**, 124710 (2014).
- [21] R. Coehoorn, C. Haas, J. Dijkstra, C. J. F. Flipse, R. A. de Groot, and A. Wold, Physical Review B **35**, 6195 (1987).
- [22] W. S. Yun, S. W. Han, S. C. Hong, I. G. Kim, and J. D. Lee, Phys. Rev. B **85**, 033305 (2012).
- [23] A. Kormányos, P. Rakyta, and G. Burkard, New J. Phys. **17**, 103006 (2015).
- [24] F. F. Fang and P. J. Stiles, Phys. Rev. **174**, 823 (1968).
- [25] T. Okamoto, K. Hosoya, S. Kawaji, and A. Yagi, Phys. Rev. Lett. **82**, 3875 (1999).
- [26] J. Zhu, H. L. Stormer, L. N. Pfeiffer, K. W. Baldwin, and K. W. West, Phys. Rev. Lett. **90**, 056805 (2003).
- [27] K. Vakili, Y. P. Shkolnikov, E. Tutuc, E. P. De Poortere, and M. Shayegan, Phys. Rev. Lett. **92**, 226401 (2004).
- [28] E. Tutuc, E. P. De Poortere, S. J. Papadakis, and M. Shayegan, Phys. Rev. Lett. **86**, 2858 (2001).
- [29] C. Attaccalite, S. Moroni, P. Gori-Giorgi, and G. B. Bachelet, Phys. Rev. Lett. **88**, 256601 (2002).
- [30] D. V. Rybkovskiy, I. C. Gerber, and M. V. Durnev, Phys. Rev. B **95**, 155406 (2017).
- [31] The $\kappa$ value for a 2D system at the interface between two dielectrics is $\kappa(t,b) = [(\epsilon_t^{\parallel}\epsilon_t^{\perp})^{1/2} + (\epsilon_b^{\parallel}\epsilon_b^{\perp})^{1/2}]/2$, where $\epsilon_{t(b)}^{\parallel/\perp}$ is the the top (bottom) relative dielectric constant with respect to the 2D plane normal. For monolayer MoSe$_2$ $\kappa = \kappa(\text{hBN}, \text{hBN})$, where $\epsilon_{\text{hBN}}^{\parallel} = 3.0$, $\epsilon_{\text{hBN}}^{\perp} = 6.9$, [32]. For bilayer MoSe$_2$ the dielectric environment of electrons in top-layer is altered by the depleted MoSe$_2$ bottom layer, for which the following average is used $\kappa = [\kappa(\text{hBN}, \text{hBN}) + \kappa(\text{hBN}, \text{MoSe}_2)]/2$, where $\epsilon_{\text{MoSe}_2}^{\perp} = 15.5$ [33], $\epsilon_{\text{MoSe}_2}^{\parallel} = 4$ [34].
- [32] R. Geick, C. H. Perry, and G. Rupprecht, Phys. Rev. **146**, 543 (1966).
- [33] A. R. Beal and H. P. Hughes, J. Phys. C: Solid State Phys. **12**, 881 (1979); Y. Li, A. Chernikov, X. Zhang, A. Rigosi, H. M. Hill, A. M. van der Zande, D. A. Chenet, E.-M. Shih, J. Hone, and T. F. Heinz, Phys. Rev. B **90**, 205422 (2014).
- [34] A. Kumar and P. K. Ahluwalia, Phys. B **407**, 4627 (2012).

# Supplemental material:
# Large effective mass and interaction-enhanced Zeeman splitting of $K$-valley electrons in MoSe$_2$

Stefano Larentis,[1] Hema C. P. Movva,[1] Babak Fallahazad,[1] Kyounghwan Kim,[1] Armand Behroozi,[1] Takashi Taniguchi,[2] Kenji Watanabe,[2] Sanjay K. Banerjee,[1] and Emanuel Tutuc[1]

[1]*Microelectronics Research Center, Department of Electrical and Computer Engineering, The University of Texas at Austin, Austin, Texas 78758, USA*
[2]*National Institute for Materials Science, 1-1 Namiki Tsukuba, Ibaraki 305-0044, Japan*
(Dated: May 11, 2018)

## MAGNETOTRANSPORT IN MONOLAYER MoSe$_2$

Figure S1 shows the longitudinal resistance ($R_{xx}$) and Hall resistance ($R_{xy}$) as function of the perpendicular magnetic field ($B$) measured in monolayer MoSe$_2$ A1 at a temperature $T = 0.3$ K, and at an electron density $n = 6.8 \times 10^{12}$ cm$^{-2}$. The data show SdH oscillations developing at $B > 9$ T, corresponding to a mobility $\mu \simeq 1100$ cm$^2$/Vs. At high $B$-fields quantum Hall states (QHSs) develop at $\nu = 8, 10, 12$; $\nu = nh/eB$, where $e$ is the electron charge, $h$ is Planck's constant. Both Fig. S1 data measured in monolayer MoSe$_2$, and Fig. 1(d) data measured in bilayer MoSe$_2$ show QHSs developing at predominantly even filling factors indicating an apparent two-fold Landau level degeneracy.

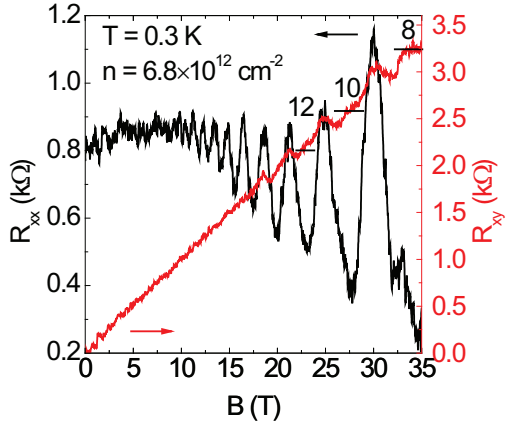

FIG. S1. $R_{xx}$ (left axis) and $R_{xy}$ (right axis) vs $B$ measured at $T = 0.3$ K and $n = 6.8 \times 10^{12}$ cm$^{-2}$ in monolayer MoSe$_2$ B1.

## FILLING FACTOR SEQUENCE TRANSVERSE ELECTRIC FIELD DEPENDANCE

In our samples a set of gate biases determines $n$, and the transverse electric-field $E = |C_{TG}V_{TG} - C_{BG}V_{BG}|/2\epsilon_0$; $C_{BG}$ ($C_{TG}$) is the bottom (top)-gate capacitance and $\epsilon_0$ is the vacuum permittivity. Figure S2 shows $R_{xx}$ vs $\nu$ measured at different $E$ ranging between $1.30 - 1.74$ V/nm, at a fixed $n = 9.2 \times 10^{12}$ cm$^{-2}$, and $T = 1.5$ K in bilayer MoSe$_2$ B2. The $R_{xx}$ minima are insensitive to the $E$-field, which suggest an $E$-field independent band and LL structure in the range of values probed here.

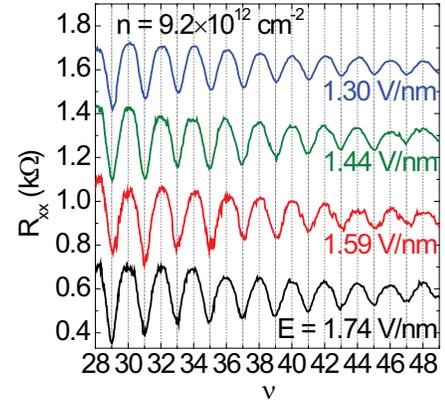

FIG. S2. $R_{xx}$ vs $\nu$ measured in bilayer MoSe$_2$ B2 at $T = 1.5$ K and $n = 9.2 \times 10^{12}$ cm$^{-2}$, and at different $E$-fields. The traces are offset for clarity.

## EFFECTIVE MASS EXTRACTION

Figure 3(a) shows the SdH oscillations $T$ dependence at $n = 4.9 \times 10^{12}$ cm$^{-2}$ for bilayer MoSe$_2$ samples B1, displaying a clear reduction in the oscillations amplitude ($\Delta R_{xx}$) as the $T$ value is increased. The $\Delta R_{xx}$ temperature dependence is proportional to the Dingle factor, $\xi/\sinh\xi$, where $\xi = 2\pi^2 k_B T/\hbar\omega_c$, $\omega_c = eB/m^*$, $m^*$ is the electron effective mass, $k_B$ is the Boltzmann constant, and $\hbar$ is the reduced Planck's constant. To extract $m^*$ we first obtain the FT amplitude spectra for the $R_{xx}$ vs $B^{-1}$ data of Fig. 3(a) [Fig. S3(a)]. A band pass filter centered around $f$, corresponding to shaded region in Fig. S3(a), is applied to eliminate other frequency components. Figure S3(b) shows $\Delta R_{xx}$ vs $B^{-1}$ data at different $T$, obtained by applying an inverse FT to the filtered spectra. The $\Delta R_{xx}$ vs $T$ data at a fixed $B$-field are fit to the Dingle factor to obtain $m^*$, as shown in Fig. S3(b) inset.

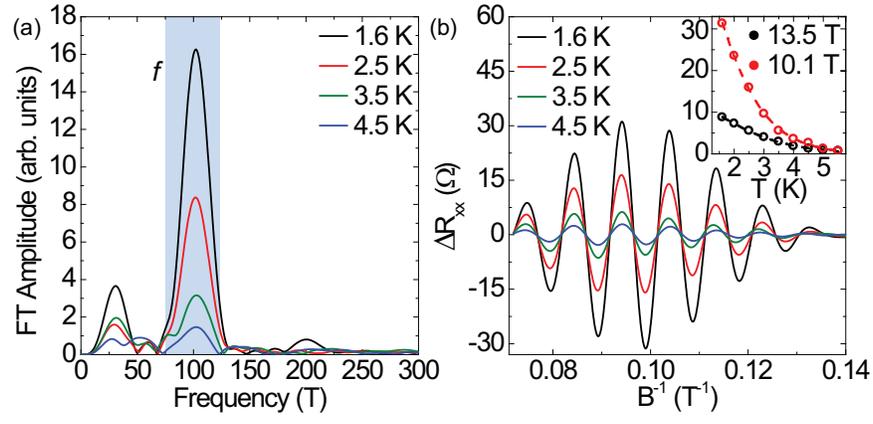

FIG. S3. (a) FT amplitude vs frequency for the $R_{xx}$ vs $B^{-1}$ data of Figure 3(a). (b) $\Delta R_{xx}$ vs $B$ calculated from the inverse FT of panel (a) data, using a bandpass filter centered around $f$ [shaded region in panel (a)]. Inset: $\Delta R_{xx}$ vs $T$ data at fixed $B = 13.5$ T, 10.1 T (symbols), and Dingle factor fit to the experimental data (dashed lines).